# Ultra-Thin Free-Standing Single Crystalline Silicon Membranes With Strain Control

Andrey Shchepetov,[1,*] Mika Prunnila,[1] Francesc Alzina,[3] Lars Schneider,[3] John Cuffe,[3,†] Hua Jiang,[4] Esko I. Kauppinen,[4] Clivia M. Sotomayor Torres[2,3] and Jouni Ahopelto[1]

[1]*VTT Technical Research Centre of Finland, P.O. Box 1000, FI-02044 VTT, Finland*
[2]*Catalan Institution for Research and Advanced Studies (ICREA) 08010 Barcelona, Spain*
[3]*Catalan Institute of Nanotechnology, Campus de la UAB, 08193 Bellaterra (Barcelona), Spain.*
[4]*NanoMaterials Group, Department of Applied Physics and Center for New Materials, Aalto University School of Science, P.O. Box 15100, FI-00076 Aalto, Finland*
[†] *Present address: MIT, Room 7-034C, 77 Massachusetts Avenue, Cambridge, MA 02139, USA.*
[*]*E-mail: andrey.shchepetov@vtt.fi*

We report on fabrication and characterization of ultra-thin suspended single crystalline flat silicon membranes with thickness down to 6 nm. We have developed a method to control the strain in the membranes by adding a strain compensating frame on the silicon membrane perimeter to avoid buckling of the released membranes. We show that by changing the properties of the frame the strain of the membrane can be tuned in controlled manner. Consequently, both the mechanical properties and the band structure can be engineered and the resulting membranes provide a unique laboratory to study low-dimensional electronic, photonic and phononic phenomena.

Silicon has been the dominant material for microelectronics since early 60's. It can be produced with very high crystalline quality, extreme purity and with well controlled electrical properties. Silicon on insulator (SOI) material has been the cornerstone in the development of state-of-the art transistors,[1] MEMS devices,[2] opto-mechanical structures,[3] photonic crystals,[4] and even lithiated anodes for batteries,[5] among others. Furthermore, thin free-standing SOI structures are ideal to investigate basic physics phenomena in low dimensional structures, such as thermal properties.[6,7,8,9] The most common way to fabricate supported single crystalline Si membranes is to release the top Si layer of SOI wafer by etching from the backside through the handle wafer and the buried oxide (BOX) layer. However, to obtain very thin membranes, the process usually requires thermal oxidation to thin down the SOI film, followed by removal of the grown oxide. Thermal processes tend to create compressive stress in the SOI film,[10] leading to buckling of the membranes after release, see Fig. 1(a). The buckling has been observed in numerous works [see, e.g., Refs. 11, 12, 13] and it can be detrimental for device applications exploiting thin free-standing structures. It also complicates experimental work, especially optical measurements, because the angle of incidence of the laser beam may not be well defined. More importantly, the strain, and, consequently the elasto-mechanical properties of the membrane, cannot be tuned in a controlled manner. Strain in a Si film can be adjusted, for example, by growing silicon epitaxially on a SiGe buffer[14] or by depositing thick SiN layers on top and bottom of the Si layer.[15] However, these approaches do not allow the fabrication of bare free-standing Si membranes with tunable strain.

In this Letter, we report on fabrication and characterization of large area ultra-thin flat suspended single crystalline Si membranes with controlled strain and thickness down to 6 nm. Our method involves insertion of a strain compensating $Si_3N_4$ frame around the Si membrane perimeter, as shown in Figs. 1(b) and 1(c). The $Si_3N_4$ layer has small tensile stress that creates a pulling force that flattens the Si membrane – similarly as a drumhead is tuned. The magnitude of the strain is related to the strain compensation ratio $R_c = w_c/w_m$, where $w_m$ is the width of the bare Si part of the membrane and $w_c$ is the width of the area of the released membrane covered with the $Si_3N_4$ film. We show that this approach not only removes the detrimental buckling but also in addition provides an elegant way to control the strain and, consequently, the strain dependent properties such as the energy band structure and elasto-mechanical properties of the lattice without degrading the crystalline quality.

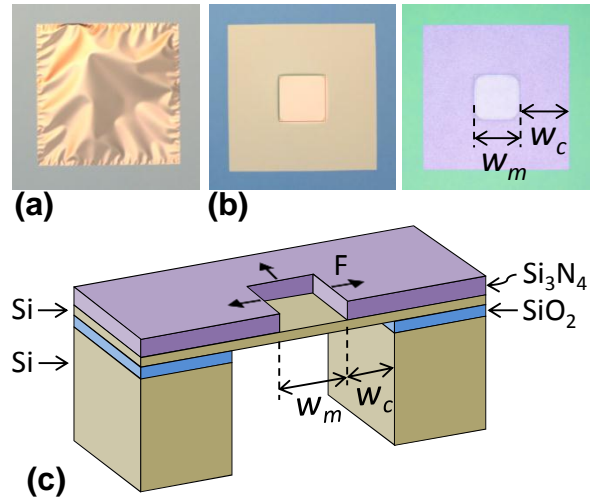

**Figure 1**. (a) Optical micrograph of a 37 nm thick free-standing Si membrane without strain compensation. (b) Optical micrographs of a strain compensated 37 and 6 nm thick Si membranes. In (a) and (b) the total width of the suspended membrane $w_s$=1000 um and in (b) $w_c$ is 350 um and $w_m$ 300 um, i.e., $R_c$=1.2. (c) Schematics of the membrane structure. $w_c$ is the distance from the edge of the membrane to the edge of the stress compensating $Si_3N_4$ layer and $w_m$ is the width of the bare Si membrane. $F$ denotes the pulling force due to the $Si_3N_4$ frame

In the fabrication commercially available bonded 150 mm SOI wafers were used. First, a 20 nm thick thermal oxide is grown on the SOI wafer and the SOI film is transferred onto a new handle wafer with the 20 nm thick oxide acting as a new buried oxide layer. The SOI film is thinned by thermal oxidation and oxide stripping to desired thickness. Then a 280 nm thick $Si_3N_4$ film under tensile stress is deposited by low pressure chemical vapor deposition (LPCVD) and patterned by UV-lithography and plasma etching, followed by deep etching from the backside through the wafer to release the membranes. The $Si_3N_4$ frame around the Si membrane provides a pulling force and prevents buckling of the membrane. This approach allows fabrication of flat free-standing single crystalline membranes with size up to several square millimeters and with thickness down to a few nanometers. The crystalline quality of the membranes was verified by high resolution transmission electron microscopy (HRTEM). The images were taken directly from the supported free-standing membranes without transferring them onto a TEM grid to reveal the true condition of the membranes. The surface profiles were measured by optical profilometry and the strain by Raman spectroscopy.

The total width of the suspended area $w_s$ ranges in these samples from 700 to 1700 um. The surface profiles of 37 nm thick membranes with and without the strain compensation, measured by an optical profilometer, are shown in Fig. 2(a). In both cases the area of the free-standing bare Si membrane is 1000 by 1000 um$^2$ and $R_c$=0.35 for the compensated membrane. The amplitude of the undulation of the non-compensated membrane is about 4 um whereas the compensated film is completely flat. All the samples with thickness ranging from 6 to 54 nm analyzed using HRTEM showed no defects and had high crystalline quality, emphasizing that the thickness or the strain does not affect the quality of the membrane. As an example a lattice image of a 9 nm thick membrane and the corresponding diffraction pattern are shown in Figs. 2(b) and 2(c). The membrane has a strain compensation ratio $R_c$=1.3, corresponding to tensile strain of about $1.2\times10^{-3}$ as will be shown below.

Raman spectroscopy is a convenient non-destructive and extremely sensitive method to measure strain in the free-standing membranes.[16] The shift of the LO phonon $\Delta\omega_{LO}$ peak at $\omega_{LO}$=520 cm$^{-1}$ wavenumber reflects the strain in the membrane. The shift of the peak can be translated to strain $\epsilon_\parallel$ as [17]

$$\Delta\omega_{LO} = \frac{1}{\omega_0}\left[q - p\frac{c_{12}}{c_{11}}\right]\epsilon_\parallel. \quad (1)$$

Here $c_{ij}$ are the stiffness constants, which for Si are $c_{11}$=1.66×10$^{11}$ and $c_{12}$=0.64×10$^{11}$.[18] The parameters $p$ and $q$ have values $p = -1.49\omega_0^2$ and $q = -1.97\omega_0^2$.[19]

Sets of 6, 27 and 54 nm thick membranes with different compensation ratio $R_c$ were characterized by Raman spectroscopy. The measurements were carried out in ambient at room temperature with excitation at 514.5 nm at different power levels to exclude the effects of potential heating of the membrane. To illustrate the effect of the compensation ratio $R_c$ and the thickness of the membrane on the induced strain, LO phonon peaks measured from sets of 54 and 6 nm thick membranes with various $R_c$ of 0.8, 1.2 and 3.5, together with a reference Si LO peak are shown in Fig. 3.

The shift of the LO phonon peak to lower energy with increasing compensation ratio indicates increasing tensile strain in the membranes. This is the strain tuning effect due to the $Si_3N_4$ frame, which can be represented in more quantitative way by transferring the peak shift into strain using Eq. (1). Fig. 4 shows the strain obtained from the measured LO phonon peak shifts for the 6 and 54, and 27 nm thick membranes as a function of the compensation ratio $R_c$. The behavior is similar for all membranes: The strain increases rapidly for small $R_c$ and then begins to saturate at larger $R_c$, showing that thicker membranes need more force to build up the strain.

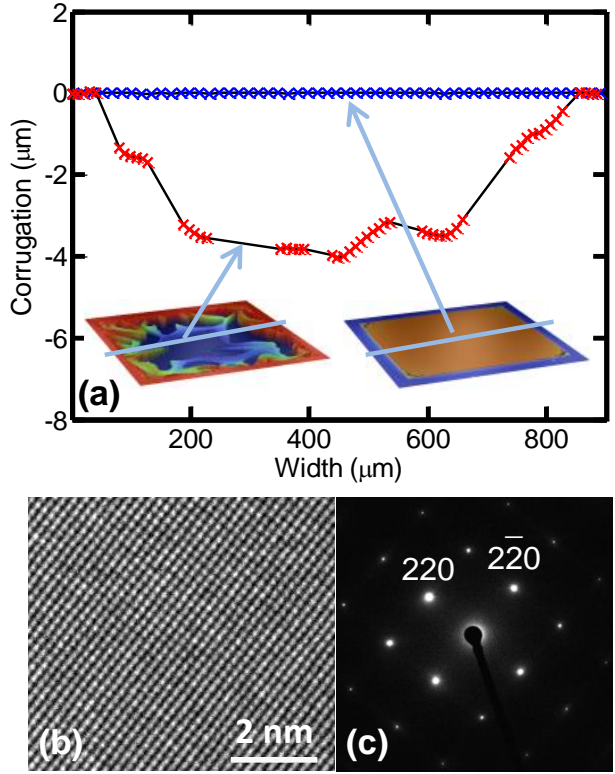

**Figure 2**. (a) Surface profiles of 37 nm thick free-standing membranes measured with optical profilometer. (Blue symbols) Surface profile of strain-compensated membrane with $w_s$=1700 um and $w_m$ 1000 um, i.e., $R_c$=0.35. (Red symbols) Non-compensated membrane with $w_s$=$w_m$=1000 um. The missing points result from a too large incident angle for detection. The solid line is a guide to the eye. In the insets lines show where the 3-dimensional profilometer images of the membranes were taken from, (left) without strain compensation and (right) strain compensated. (b) High resolution TEM image of a 9 nm thick Si membrane with $w_c$=400 um and $w_m$=300 um, i.e., the strain compensation ratio $R_c$=1.3. (c) Diffraction pattern of the same membrane.

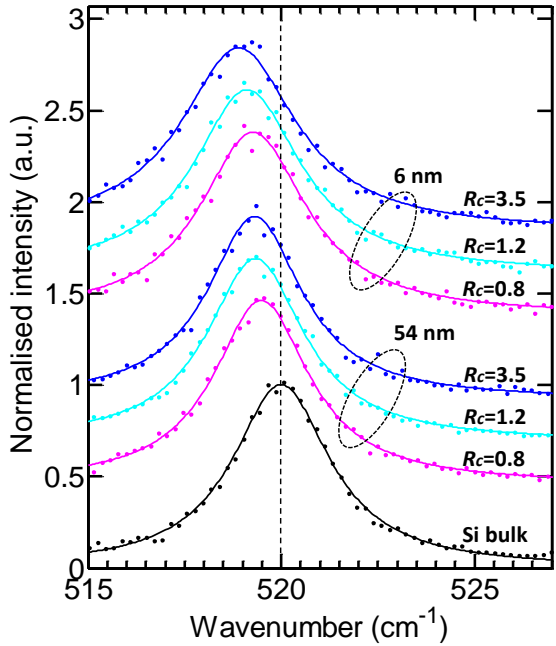

**Figure 3.** LO peaks measured by Raman scattering from 6 and 54 nm thick free-standing Si membranes with the strain compensation ratio $R_c$ of 0.8, 1.2 and 3.5. The bottom curve is a reference spectrum measured from the bulk part of the Si wafer, defining the reference peak position $\omega_{LO} = \omega_0$ in Eq. (1).

The strain has an effect on many properties of the membranes, such as the mechanical resonance frequencies, Q-value and electronic band structure. Silicon has six equivalent conduction band energy minima and the biaxial strain lifts the degeneracy of these valleys. The energy of the two out-of-plane valleys decreases compared to the four in-plane valleys, as shown in the inset to Fig. 4, leading to changes in their relative population,. The energy splitting $\Delta E_c$ between the in-plane and out-of-plane valleys of the silicon conduction band can be written as[20]

$$\Delta E_c = \Xi_u \left[1 + 2\frac{c_{12}}{c_{11}}\right]\varepsilon_{\parallel}, \quad (2)$$

where $\Xi_u$ = 9.0 eV is the uniaxial deformation potential.[18] The strain induced energy splitting is shown in Fig. 4 in the right axis. At large $R_c$ the splitting is of the order of the thermal energy at room temperature. This splitting has an effect, for example, on carrier mobility[21] and on thermal electron-phonon coupling[22] in the membrane. The strain affects also the valence bands and the band gap $E_g$. There is no simple analytical formula for the energy shift of valence bands and here the results of Ref. 23 are used, giving the dependence of the band gap energy $\Delta E_g$ for small strains as $\Delta E_g \approx -19.15 \times \varepsilon_{\parallel}$ eV. The change in the band gap $\Delta E_g$ as a function of $R_c$ is plotted in the second right y-axis in Fig. 4. The maximum $\Delta E_g$ in the current membranes is ~30 meV which can be detected by absorption or photoluminescence measurements.

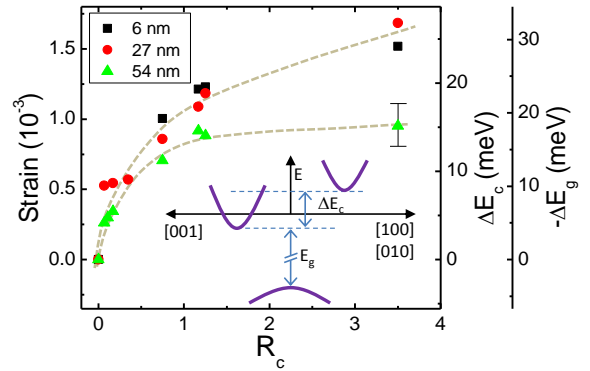

**Figure 4.** Strain as a function of strain compensation ratio $R_c$ (left y-axis) in Si membranes with thickness of 6 nm, 27 nm and 54 nm. (First right y-axis) Corresponding energy splitting $\Delta E_c$ between the in-plane and out-of-plane conduction band minima in Si. (Second right y-axis) Change of the band gap energy $\Delta E_g$. The splitting of the out-of-plane [001] and in-plane [100]/[010] conduction band minima and the energy band gap are shown schematically in the inset. The dashed curves are guides for the eye and the error bar is the same for all data points.

The LPCVD silicon nitride used in these experiments for strain compensation has a relatively low tensile stress of about 300 MPa. By varying the growth parameters the stress can be varied between 0 and 1000 MPa, providing a mean to increase substantially the strain in the membranes, and consequently, to tune the properties of the membranes. The thinnest membranes fabricated here are 6 nm thick, corresponding roughly to 20 atomic layers. The membranes are still very robust and preserve their crystallinity surprisingly well, even under the tensile strain. Thermal oxidation is an extremely well controlled process and the interface between silicon and the thermal oxide is very sharp, of the order of one atomic layer. We expect that free-standing membranes of thickness of 1-2 nm can be realized, having only a few atomic layers remaining. In these membranes reconstruction of the lattice will probably begin to play a role and the theoretically predicted effects in very small crystalline structures[24] can be investigated experimentally. The role of the native oxide is not clear yet regarding the properties of the thinnest membranes. Native silicon oxide is eventually formed on silicon surfaces in ambient conditions. The thickness of the oxide layer is of the order of 1 nm but the volume is about twice of the consumed silicon, leading to inhomogeneous thickness of the oxide film and, potentially, local straining of the underlying lattice. Furthermore, the presumed reconstruction of the lattice may alter the physical properties and chemical activity of silicon.

In summary, we report a method to fabricate free-standing Si membranes with thickness ranging down to sub-10 nm and the strain accurately controlled. The strain enables the manipulation of the band structure of Si, permitting engineering of, e.g., electron-phonon and phonon-phonon coupling in 2-dimensional system, and investigation of the properties of low dimensional systems by optical means due to the large-area of the flat membranes. The fabricated membranes are robust and realization of even thinner membranes with high strain using the demonstrated approach is envisaged.

The authors acknowledge the financial support from the FP7 projects NANOFUNCTION (grant nr. 257375) and NANOPOWER (grant nr.ICT-2010-256959). LS, JC, FA and CMST acknowledge the support of the Spanish projects nanoTHERM (grant nr. CSD2010-0044) and ACPHIN (FIS2009-10150). MP and AS acknowledge funding from the Academy of Finland (grant nr. 252598).